# Machine Learning Application in the Life Time of Materials

## Xiaojiao Yu

**Abstract:**

Materials design and development typically takes several decades from the initial discovery to commercialization with the traditional trial and error development approach. With the accumulation of data from both experimental and computational results, data based machine learning becomes an emerging field in materials discovery, design and property prediction. This manuscript reviews the history of materials science as a disciplinary the most common machine learning method used in materials science, and specifically how they are used in materials discovery, design, synthesis and even failure detection and analysis after materials are deployed in real application. Finally, the limitations of machine learning for application in materials science and challenges in this emerging field is discussed.

1. Introduction

Materials science has a long history that can date back to the Bronze age [1]. However, only until the 16th century, first book on metallurgy was published, marking the beginning of systematic studies in materials science [2]. Researches in materials science were purely empirical until theoretical models were developed. With the advent of computers in the last century, numerical methods to solve theoretical models became available, ranging from DFT (density functional theory) based quantum mechanical modeling of electronic structure for optoelectronic properties calculation, to continuum based finite element modeling for mechanical properties [3-4]. Multiscale modeling that bridge various time and spatial scales were also developed in the materials science to better simulate the real complex system [5]. Even so, it takes several decades from materials discovery to development and commercialization [6-7]. Even though physical modeling can reduce the amount of time by guiding experiment work. The limitation is also obvious. DFT are only used for functional materials optoelectronic property calculation, and that is only limited to materials without defect [8]. The assumption itself is far off from reality. New concept such as multiscale modeling is still far away from large scale real industrial application. Traditional ways of materials development are impeding the progress in this field and relevant technological industry.

With the large amount of complex data generated by experiment, especially simulation results from both published and archived data including materials property value, processing conditions, and microstructural images, analyzing them all becoming increasingly challenging for researchers. Inspired by the human genome initiative, Obama Government launched a Materials Genome Initiative hoping to reduce current materials development time to half [9]. With the increase of computing power and the development of machine learning algorithms, materials informatics has increasingly become another paradigm in the field.

Researchers are already using machine learning method for materials property prediction and discovery. Machine learning forward model are used for materials property prediction after trained on data from experiments and physical simulations. Bhadeshia et al. applied neural network (NN) technique to model creep property and phase structure in steel [10-11]. Crystal structure prediction is another area of study for machine learning thanks to the large amount of structural data in crystallographic database. K-nearest-neighbor's method was used to identify materials' structure type based on its neighbors' structure types [12-13]. Machine learning is also applied for materials discovery by searching compositional, structural space for desired properties, which is essentially solving a constrained optimization problem. Baerns et al. was able to find an effective multicomponent catalyst for low-temperature oxidation of low-concentration propane with a genetic algorithm and neural network [14].

There are a few reviews on machine learning application in materials science already. Dane Morgan and Gerbrand Ceder reviewed the data mining methods in materials development [15]. Tim Mueller, Aaron Gilad Kusne, and Rampi Ramprasad also reviewed the progress and application of machine learning in materials science, more specifically in phase diagram, crystal structural and property prediction [16]. However, their reviews are mostly based on applications in fundamental of materials science. Here, we are taking a more practical approach of reviewing machine learning application in material design, development and stages after deployment. We first discuss data problems specifically in materials science. Then, machine learning concept and most widely used methods are introduced. Up-to-date reviews on machine leaning application in materials discovery, design, development, deployment and recall is conducted. The relation between data driven research and traditional experimental and physical modeling is discussed afterwards. Finally, challenges and future endeavors of machine learning based materials science research is pointed out for researchers in this niche area.

## 2.1 Data Problem in Materials Science

The successful application of informatics in biology, astronomy and business has inspired similar application in materials science. However, materials science differs from other subjects due to its unique characteristics. Some researchers are debating whether there is a big data problem in materials science, after all the size of materials data is nothing comparable to biology data. The largest existing database based on experimental results from materials has $5 \times 10^5$ data records [17]. However, the rapid progress in computational science and microscopy techniques is resulting in enormous amounts of output data [18]. Furthermore, Materials science data tends to be complex and heterogeneous in terms of their sources and types ranging from discrete numerical values to qualitative descriptions of materials behavior and imaging data [19]. Data in materials science also exhibit the Veracity characteristics of big data problem, by that we acknowledge the practical reality of data missing and uncertainties with the data [19]. According to the 4V (volume, variety, velocity, veracity) characteristics of big data, materials science does have a big data problem [19]. With the emergence of this big data in materials science, how to extract hidden information from the complex data and interpret resulted information is becoming increasingly important for materials design and development.

## 2.2 Machine Learning Methods

Machine learning, a branch of artificial intelligence, is about computer learning from existing data without being explicitly programmed and make predictions on new data by building a model from input samples. Depending on the assigned task, machine learning can be classified into three categories: supervised learning, machine learning algorithms are trained with a set of input value and labeled output value first, then they are used to predict output values for corresponding unseen input values; unsupervised learning, where there is no labelled output value for training data and machine learning algorithm is used to discover patterns in the input value; reinforcement learning (program interact with environment dynamically to maximize accumulated rewards). Reinforcement learning is not used in materials science field; hence it is not introduced in detail in this manuscript. Supervised learning can either be a classification problem or a regression problem depends on the whether the output value is discrete or continuous.

## 2.3 Method Workflow

Machine learning method typically comprise several steps including raw data collection, data preprocessing (filling in missing data, handling outliers, data transformation), feature engineering for feature selection and extraction (principle component analysis), model selection, training, validations and testing. A detailed workflow is presented in Fig 1. To select the best algorithm for a particular task, model evaluation is important. Different algorithms are evaluated with different metrics. For instance, a classifier's evaluation metrics include confusion matrix, AUC (area under curve), precision recall, F measure, Kolomogorov Smirnov chart (K-S). Confusion matrix is a 2X2 matrix with four elements: true positive (TP), true negative(TN), false positive (FP), false negative(FN) [20]. Other accuracy measures are sensitivity (True Positive Rate=$\frac{TP}{TP+FN}$), specificity (True negative rate=$\frac{TN}{TN+FP}$). AUC is the area under ROC curve, which consider the relation between sensitivity and specificity. The greater the area under the curve, the more accurate is the model. Precision is $\frac{TP}{TP+FP}$, recall is the true positive rate defined as above. Precision-recall shows the fraction of predictions that are false positive [21]. F measure is also a measure of the model accuracy and is defined as the weighted harmonic mean of the precision and recall of the test. F is the balance between precision and recall [22]. K-S evaluate how the model separates between the positive and negative distributions. Higher KS value means better separation [23].

For regression algorithms, evaluation metric includes mean absolute error, (root) mean squared error (RMSE = $\frac{\sqrt{\sum_{i=1}^{N}(y_i-\hat{y}_i)^2}}{N}$), coefficient of determination ($R^2$). $R^2 = \frac{regression\ variation}{total\ variation} = \frac{\sum_{i=1}^{N}(\hat{Y}_i-\bar{Y})^2}{\sum_{i=1}^{N}(Y-\bar{Y})^2}$ measures the percent of total variability that is explained by the regression model [24].

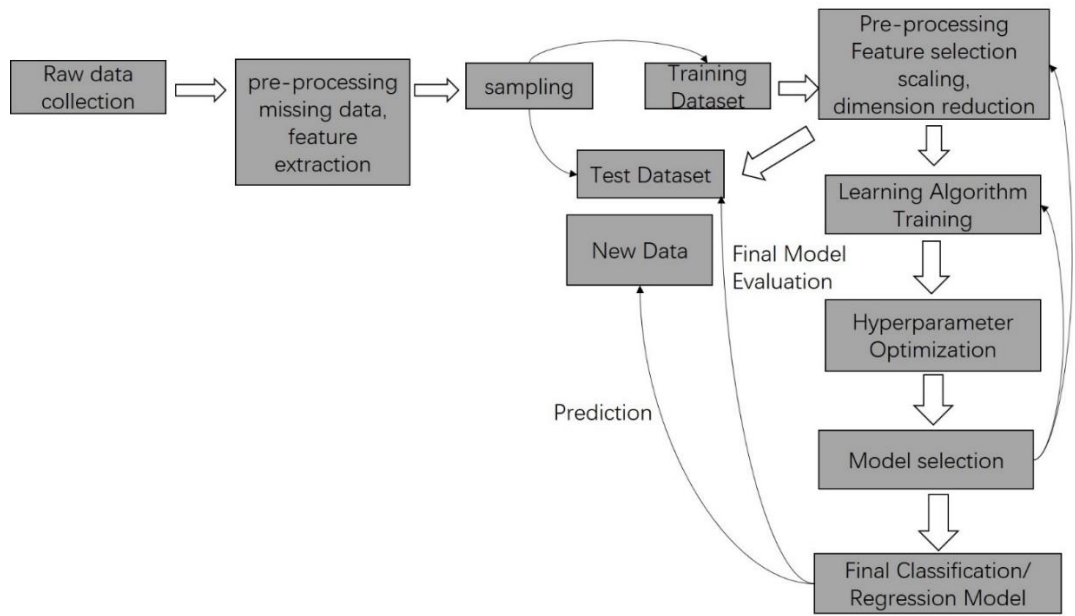

Fig. 1. Flowchart of a typical machine learning method

## 2.4 Method Comparison

Some of the most common machine learning algorithms are SVM (support vector machine), ANN (artificial neural network), logistic regression, decision trees. Support vector machine algorithms are used to find the hyperplane that separate different classes with highest margin [25]. The advantage of SVM is that the solution is global and unique. Computation complexity of SVM does not depend on the dimension of the input space and is less prone to overfitting [26-27]. However, SVM does not work well on unbalanced data [26]. Artificial neural network is inspired by biological brain, where artificial neurons are connected to mimic the connection of neurons in the brain [28]. Multiple hidden layers and neurons can add to the complexity of the neuron network architecture. The strength of ANN is that they are flexible and can represent any nonlinear and linear function. However, it needs large amount of training data and is prone to overfitting. Hyperparameter tuning is tedious and troublesome for ANN. Decision tree is another commonly used basis classification algorithm, which comprises a root node, internal node, branch, leaf node, and depth [29]. Decision tree progressively splits the tested data based on input feature value, decision process follows the branch, which is the collection between an internal node and its parent node, until it reaches a leaf node [30]. Ensemble methods such as random forest and adaboost which are based on constructing a large number of trees with bootstrap samples and iteratively build an ensemble of weak learners, in an attempt to generate a strong overall model. Ensemble methods usually perform better than basic machine learning algorithms in terms of reducing variance and bias [31].

### 3.1 Machine Learning Application in Materials Discovery and Design

An important concept in materials science field is structural-property-performance relationship [32]. Developing materials that meet the required performance and property goes back to control processing conditions, structural and compositions of the materials. Hence, understanding how processing condition, structural and compositions affect materials property and performance is the first step towards materials design. Traditionally, controlled experiments are conducted to isolate the effect of one variable. However, variables often are correlated with each other. It is infeasible to isolate some variable for experimental testing [33]. Data mining can help revealing hidden relations between large amount of materials parameters, processing conditions and their relations with dependent materials properties [33]. Traditional ways of materials development can be disrupted and reshaped by making the use of available data.

### 3.1.1 Materials Property Prediction

Materials design first of all requires understanding of how desired properties such as materials' yield strength, toughness, ultimate tensile strength, and fatigue life etc. are affected by intrinsic microstructure, chemical composition, crystal structure, and external processing, loading conditions and temperatures. Machine learning algorithm can derive the quantitative relation between the independent and dependent variables and hence make prediction with enough training data when physical model does not exist or is too complicated to apply [33].

Neural network algorithm has been used in ferritic steel welds toughness prediction due to their ability to handle complex models [33]. Toughness was studied as a function of chemical composition, microstructure, welding process and testing temperature. Their influence on toughness was shown in Fig 2. The interaction between different variables can also be predicted with neural network algorithm as shown in Fig 3. The cross of the two toughness curves as a function of temperature and manganese compositions indicates at higher temperatures the influence of manganese on toughness was not only reduced but also negative.

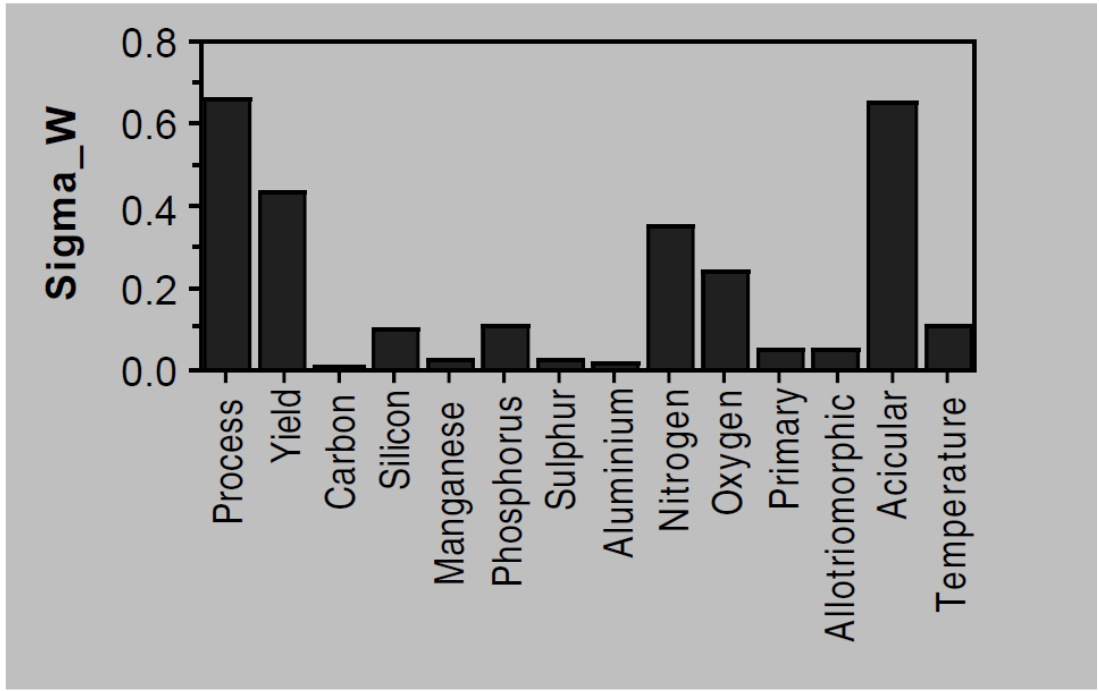

Fig 2. Bar chart showing a measure of the model-perceived significance of each of the input variable in influencing toughness. [33]

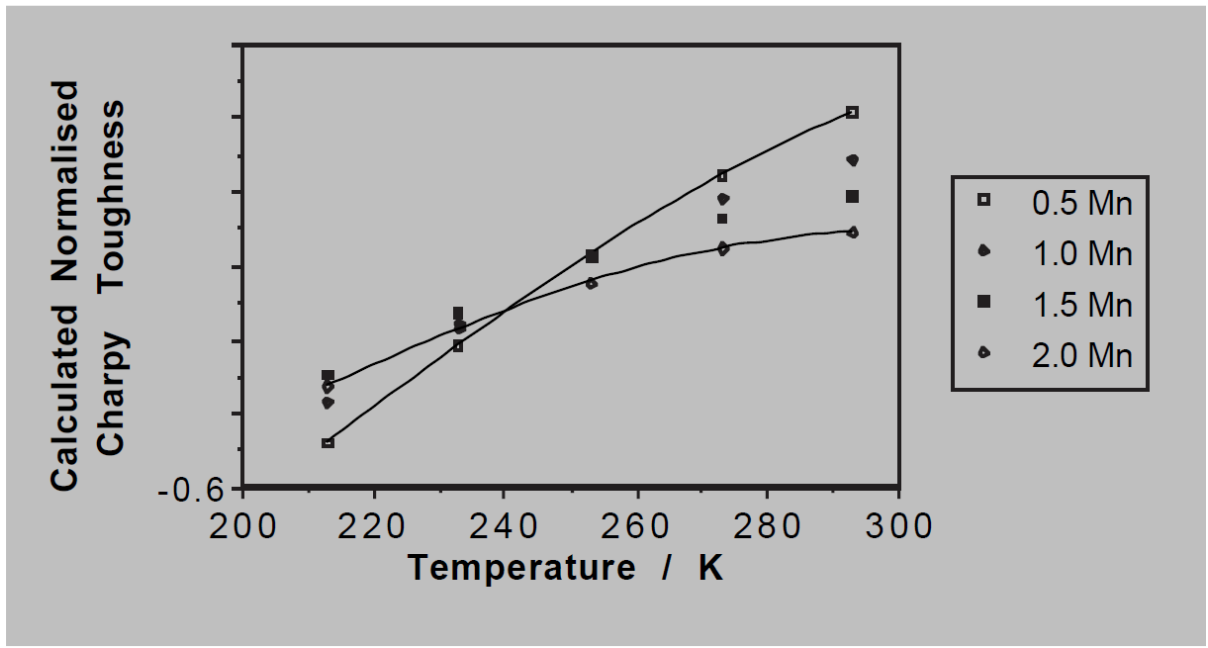

Fig 3. Variation in the normalized toughness as a function of the manganese concentration and the test temperature. [33]

ANN can also be used to predict constitutive relations. For instance, the constitutive flow behavior of 42CrMo Steel is predicted with strain, log strain rate and temperature as input and flow stress as output. Predicted results show good correlation with experimental value, indicating excellent capacity of the developed model in predicting flow stress, Fig 4 [34]. Austenite Stainless Steel grade 304L and 316L ultimate tensile strength, yield strength, tensile elongation rate, strain hardening exponent and strength coefficient were also able to be predicted by ANN with a function of temperature and strain rate. The optimum architecture is [2-6-5] for ASS 304L and [2-17-5] for ASS 316L using feed forward back propagation learning. Model accuracy is verified with correlation coefficient, average absolute error and its standard deviation [35].

Fatigue properties has always been among the most difficult ones to predict due to the high cost and long time for fatigue testing and the prevalence of structural failure caused by fatigue [36-37]. Existing physical models are either lacking of generality or fail to give quantitative indications [38]. Agrawal et al. predicted the fatigue strength of steel using data from the Japan National Institute of Materials Science (NIMS) MatNavi database [39-40]. They used 12 regression-based predictive model, among them, neural network, decision tree and multivariate polynomial regression were able to achieve a high $R^2$ value of >0.97.

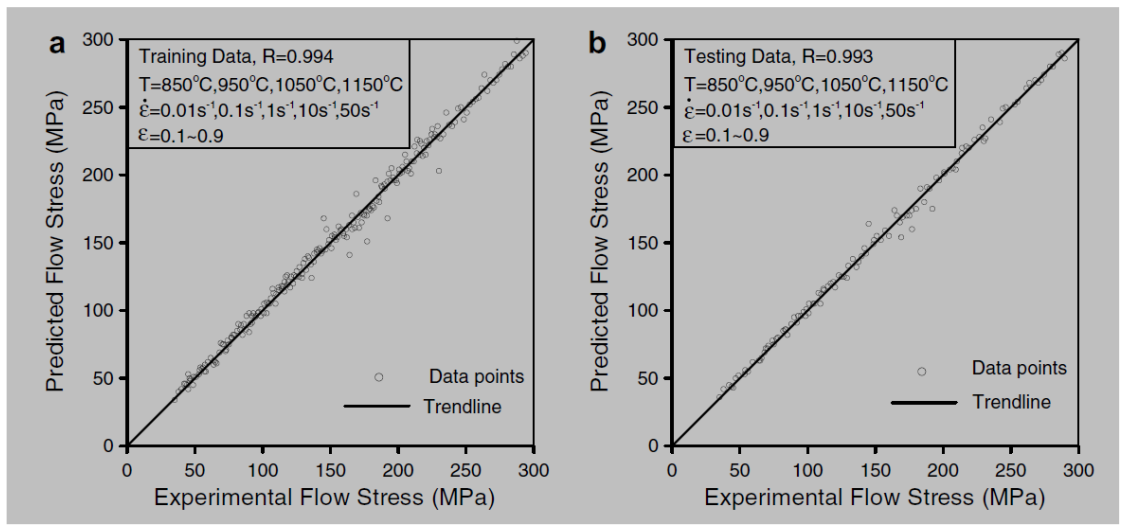

Fig 4. Comparison between experimental value and predicted flow stress of 42CrMo steel using BP ANN. [34] (a) Predicted training data (b) Predicted testing data

### 3.1.2 Inversed Design of Materials

Understanding how mechanical properties are influenced by materials internal and external factors help reducing searching space in the inversed materials design task. However, the inverse problem is more challenging because of the possibility of multiple solutions and the enormous structural dimension [41]. Machine learning application has shown promise in inversed materials discovery and design by reducing searching path and searching region. Ruoqian Liu et al. developed a machine learning method for the inverse design of Fe-Ge alloy microstructure with enhanced elastic, plastic and magnetostrictive properties [41]. A systematic approach consisting of random data generation, feature selection, and classification was developed. Firstly, features that can quantitatively describe microstructures and properties were developed. Then, randomly generated structural and properties pairs were simulated to form the most desired and least desired classes. Two crucial steps, search path refinement and search space reduction are conducted prior to the actual searching to find the most efficient orders of features in search and the most promising search regions of features. This method was validated with five design problems, which involves identification of microstructures that satisfy both linear and nonlinear property constraints. This framework shows supremacy comparing with traditional optimization methods in reducing as much as 80% of running time and achieving optimality that would not be attained.

**3.2 Machine learning application in materials processing and synthesis**

Design of materials can be facilitated with the data driven machine learning approach, however the commercialization of materials is still impeded by the availability to synthesize them. To disrupt the trial and error synthesis methods, Olivetti group in MIT is working on creating a predictive synthesis system for advanced materials processing. They are building a curated database of solid state materials and their synthesis methods compiled from thousands of materials synthesis journal articles. The database also contains algorithms developed through machine learning approaches, which are capable of predicting synthesis routes for novel materials based on chemical formulae and other known physical input data.

- Even failed experiments can be used by the machine learning algorithm for materials discovery and synthesis which truly shows the power of data mining and machine learning. After all, only a small amount of information is published in the research work, most of the data are archived and not been used to its full potential. Paul Raccuglia, et al. trained a machine learning model based on failed hydrothermal syntheses data to predict reaction outcomes under different conditions such as temperature, concentration, reactant quantity and acidity [42]. The model was validated and tested with previously untested data and shown better performance than human researchers who have 10 years' experience. It was able to predict conditions for new organically templated inorganic product formation with a success rate of 89%.

**3.3 Machine learning application in microstructure recognition and failure analysis**

Microstructure damage and failure pre-detection is another area that machine learning find its applications. Traditionally, materials scientist examines the SEM, OPM images of samples for failure analysis similar to medical doctors analyze X-Ray images of patients. With the increasing penetration of machine learning methods in medical imaging analysis, the same kind of application in materials imaging is expect to happen as well [43].

In fact, there are already reports on machine learning and computer vision researches on materials microstructure automatic recognition. Aritra et al. applied computer vision methods to identify images that contain dendritic morphology and then classify whether the dendrites are along the longitudinal direction or traverse direction if they do exist in the image. To extract features and reduce feature dimensions, they used visual bag of words, texture and shape statistics, and pre-trained convolutional neural network. Classification was conducted using support vector machine, nearest neighbors and random forest models [44]. It was shown that pre-trained convolutional neural network performs best in terms of micrograph recognition and feature extraction, which confirmed with other reports [45-46]. Classification methods were able to reach great accuracy for both task. Another example is the automatic measurement of ferrite volume fraction from the ferrite-austenite binary phase structures based on GPF (Graph Processing Framework) algorithm developed by Hafiz Muhammad Tanveer, et al [47].

Machine learning algorithm can also be used in failure detection by examining microstructure images. Matthias Demant et al. introduced an enhanced machine learning algorithm for crack detection in photoluminescence (PL) images of as-cut wafers. The detection algorithm is based on a classification of cracks due to the comparison of the crack descriptions with previous trained crack data. Crack centers are identified by detecting features appearing as star or line-like structure. Grain boundary information is extracted from additional images in the visible range to avoid false detections. Support vector machine is used to train labelled data for crack and non-crack structures classification [48]. The algorithm is able to achieve a high precision of 91.1% and sensitivity of 80.4% for crack length greater than 3 mm. Elaheh Rabiei et al. developed a dynamic Bayesian network (DBN) based on the variation of modulus of elasticity to estimate damages from a prognostic approach when crack is not observable yet. Various sources of information were taken into account to reduce uncertainties. DBN was applied to relate the variables and their causal or correlation relationship. Degradation model parameters are learned with joint particle filtering technique. Support vector regression models was applied to define unknown nonparametric and nonlinear correlation between the input variables. More precise damage estimation and crack initiation prediction in a metallic alloy under fatigue was confirmed by experimental observations [49]. This method is different from traditional empirical damage models (Paris law) since direct damage indicators such as crack is not required to predict damage stage. Thus, underling damages can be monitored at an earlier stage. It is easy to imagine manufacturing companies such as GE can monitor their jet engine data to predict whether it needs inspection or maintenance.

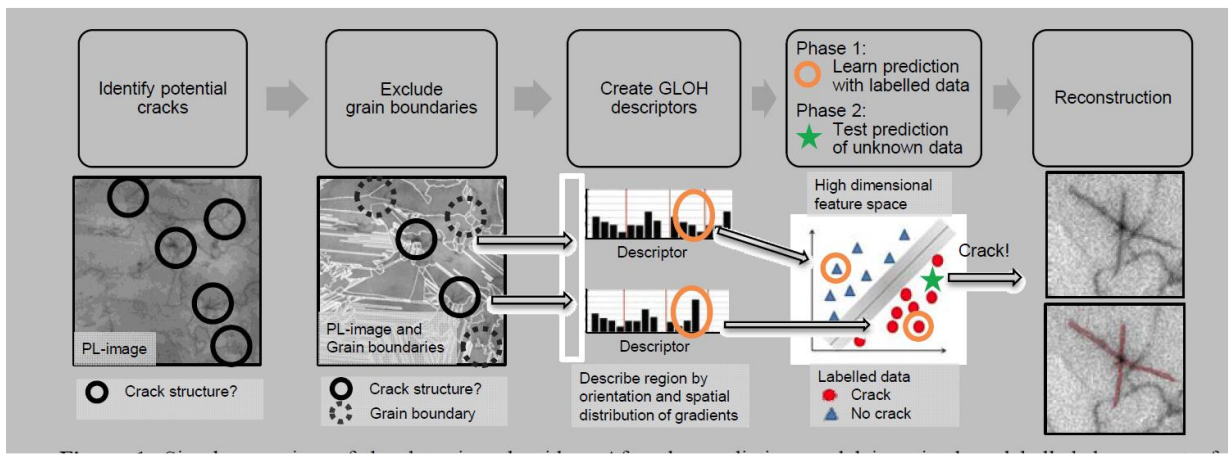

Fig. 5. Overview of the crack detection algorithm [48].

### 4. limitations of machine learning in materials science applications

Although machine learning has been widely used in a lot of fields and increasingly been used in materials science, machine learning is by no means a panacea. Without understanding its limitations and blindly apply it to every possible area can lead to wrongful predictions and a waste of time and effort. First of all, machine learning system are opaque, making them very hard to debug. Machine learning prediction heavily relies on training data. Machine learning often have overfitting or overfitting problems that needs to be concerned when taking their prediction results into consideration. Input data quality needs to be ensured. Interpolation and extrapolation can lead to problems when training data is not sufficient in the interpolated or extrapolated regime or when training data is noisy. Hence, error bar prediction is needed for evaluating prediction accuracy. Machine learning does not explain the results from the physics point of view. Materials scientists often are interested in understanding the mechanism of certain phenomena. Machine learning cannot elucidate the mechanism since it works on data driven model training and prediction. Interpretation of the machine learning results needs domain knowledge. Without understanding the underline physics, nonsense predictions can't be recognized. Even in the process of feature selection, a good understanding of the causal relationship between these variable and dependent properties can be helpful for selecting most effective features and build less complicated models.

Machine learning is also inseparable from experiment and physical simulation. It is typically used as a supplemental tool for materials discovery, design and property prediction. Machine learning training data are either from experimental results or physical simulation results [50]. Machine learning models also rely on experiments or simulations for validation. To advance this field, people from different discipline, both experimentalist and computational scientist, should collaborate on data collection, storage and curation. Interdisciplinary researchers need to be trained to understand both materials science and machine learning [51].

-